\documentclass[aps,groupedaddress,amsmath,amssymb]{revtex4-1}
\usepackage{graphicx}  
\usepackage{dcolumn}   
\usepackage{bm}        
\usepackage{verbatim}   

\def\be{\begin{equation}}
\def\ee{\end{equation}}
\def\ba{\begin{eqnarray}}
\def\ea{\end{eqnarray}}

\newcommand{\bsb}{\boldsymbol}

\def\q{{\bsb q}}

\usepackage{colortbl}
\definecolor{summersky}{cmyk}{0.71,0.33,0,0.14}
\definecolor{flamingo}{cmyk}{0,0.51,0.71,0.14}
\definecolor{rp}{cmyk}{0.2, 1, 0.6, 0}
\definecolor{pacificblue}{cmyk}{0.95,0.3,0, 0.19}
\definecolor{gray60}{cmyk}{0.4,0.4,0,0.8}
\definecolor{green94}{cmyk}{.94,0,1.00,0}
\definecolor{green80}{cmyk}{.80,0,.90,0}
\begin{document}
	\title{Primordial black holes in linear and non-linear regimes}
	
\author{Alireza Allahyari}
\affiliation{Department of Physics, Sharif University of Technology,
	Tehran, Iran }
\affiliation{ School of Astronomy, Institute for Research in Fundamental Sciences (IPM), P. O. Box 19395-5531, Tehran, Iran }
\email{allahyari@physics.sharif.edu}

\author{Javad T. Firouzjaee}
\affiliation{ School of Astronomy, Institute for Research in Fundamental Sciences (IPM), P. O. Box 19395-5531, Tehran, Iran }
\email{j.taghizadeh.f@ipm.ir}

\author{Ali Akbar Abolhasani}
\affiliation{Department of Physics, Sharif University of Technology,
	Tehran, Iran }
\affiliation{ School of Physics, Institute for Research in Fundamental Sciences (IPM), Tehran, Iran }

	\begin{abstract}
	 Using the concept of apparent horizon for dynamical black holes, we revisit the formation of primordial black holes (PBH) in the early universe for both linear and non-linear regimes. First, we develop the perturbation theory for spherically symmetric spacetimes to study the formation of spherical PBHs  in linear regime and  we fix two gauges. We also introduce a well defined gauge invariant quantity for  the expansion. Using this quantity, we argue that PBHs do not form in the linear regime. Finally, we study the non-linear regime. We adopt the spherical collapse picture by taking a closed FRW model in the radiation dominated era to investigate  PBH formation. Taking the initial condition of the
	 spherical collapse from the linear theory of perturbations, we allow for both density and velocity perturbations. Our model gives a constraint on the velocity perturbation. This model also predicts that the apparent horizon of PBHs  forms when $\delta > 3$. Applying the sound horizon constraint, we have shown the threshold value of density perturbations at horizon re-entry must be larger than $\delta _{th} >  0.7$ to overcome the pressure gradients.
	\end{abstract}
	%
	%
	
	\maketitle
	
	\tableofcontents

\section{Introduction}
In the standard model of cosmology, the astrophysical objects are originated from the early universe quantum fluctuations which became classical as they were stretched to superhorizon scales in an exponentially expanding period. If the density perturbations exceed some threshold value,  primordial black holes (PBHs) might form. Many numerical investigations have been done to study the threshold value of the density perturbations. \cite{Carr:1975qj,Niemeyer:1999ak,Musco:2004ak,Young:2014ana}. The threshold in terms of metric perturbations is investigated in \cite{Shibata:1999zs}. 
 See Reference \cite{new-thre} for recent numerical studies and references therein. Many studies are focused on these black  holes since these black holes are the most important candidate for the Hawking radiation of
 black holes \cite{Page:1976wx}. Moreover, the dark nature of these black holes could make them one of the nominated objects for some fraction of the dark matter \cite{Carr:2009jm}. \\

The evolving nature of the universe reveals that the primordial black holes are classified as  dynamical black holes \cite{Ashtekar:2004cn}. These black holes  belong to the larger class of cosmological black holes which evolve in the FRW background. In contrast to the stationary black hole cases, we need a more subtle approach to define mass \cite{Firouzjaee:2010ia}, horizon \cite{cosmological black hole} and Hawking radiation scenario \cite{tunnelingbh,firouzjaee-ellis14}  for these black holes. Located in the cosmological expanding background, the cosmological black holes have special properties at late times \cite{firouzjaee-penn}. As stated, the other important feature of the black holes is the black hole horizon. The event horizon is often attributed as the black hole boundary. But the global nature of event horizon is not appropriate for evolving black holes which are studied in the numerical relativity. Consequently, the apparent horizon is used to distinguish the black hole boundary in general cases. The apparent horizon is not a global quantity and it is quantified by the expansion of the null geodesics.\\

Early works by Carr et. al \cite{Carr:1975qj} put a lower and an upper threshold for density contrast  for PBH formation. But recently Kopp, Hofmann, and
Weller \cite{hofman-2011} have shown that the maximum value $\delta_{max}$ is
not directly related to the separate universe but to the
geometry of the over-dense region. Moreover, even when the density perturbation is greater than the threshold value, this does not guarantee that PBHs form. The necessary and essential condition is formation of the black hole's apparent horizon which means that the trapped surface forms.  It is not explicitly mentioned if PBH form in the linear regimes and superhorizon scales. The other  point  not taken into account in analytical models is matching to linear initial fluctuations. \\

In this work we revisit the problem of PBH formation in both linear and non-linear regimes. In the linear regime we first revisit the PBH formation  with emphasis on gauge invariance of its apparent horizon. To do so, we first develop the linear perturbation theory for  spherically symmetric perturbations to study the primordial black holes formation in  Section II. We use this gauge invariant quantity to argue that PBHs do not form in linear regime. 
 Section III is devoted to the PBH formation in non-linear regime with emphasis on initial conditions. We employ a spherical collapse model to find the threshold value of PBH formation. Our model initial condition is set by perturbation theory at early times. We show that when a black-hole forms, the density contrast $\delta$  has the universal value $3$. We then estimate the threshold value of the density contrast at horizon re-entry necessary for PBH formation.  Finally, we conclude in section IV.

\section{Pbh formation in linear regime}
In the standard model of cosmology it is known that all structures form from the early universe density perturbations which also leave some imprints on the CMB as fluctuations on the average temperature we measure. The wavelength of the perturbations were greater than the Hubble horizon at that time. When these linear perturbations enter the Hubble horizon they might start growing and eventually collapse into non-linear structures. In this section, we  consider the possibility that primordial black holes form in the linear regime. Since we study the formation of PBHs with  spherical symmetry, we will start off by studying  spherically symmetric perturbations. \\

\subsection{Perturbation theory with a symmetry}
To study the precise behavior of the PBHs, we need to do simulations in numerical relativity which is complicated except for simplified models. But analytical calculations may provide physical insights into the very nature of black hole formation. Considering a black hole with spherical symmetry is one of the most common assumptions that allows us to define many analytical quantities such as mass, horizon and light cone dynamics \cite{cosmological black hole}. The general perturbation theory on spherically symmetric backgrounds is presented in appendix B and Einstein equations in appendix C. In this section we assume spherical perturbations.\\

Suppose,  we have a background which is perturbed as we keep the spherical symmetry.  This means that the Lie derivative of the perturbed metric vanishes with respect to its killing vectors. To first order we have
\begin{equation}
{\pounds}_{\bar{ \varepsilon}+\delta\varepsilon}(\bar{g}+\delta g)=0,
\end{equation}
where $\varepsilon $s are the killing vectors. If we make a gauge transformation, the perturbed metric will transform as
\begin{equation}
\tilde{\delta g} =\delta g-{\pounds}_{\Delta x}\bar{g}.
\end{equation}
The killing vectors also transform as
\begin{equation}
\tilde{\delta \varepsilon}=\delta\varepsilon-{\pounds}_{\Delta x}\bar{\varepsilon}. 
\end{equation}
To have spherical symmetry we require that
\begin{equation}
{\pounds}_{\bar{\varepsilon}+\tilde{\delta \varepsilon}}(\bar{g}+\delta g-{\pounds}_{\Delta x}\bar{g})=0.
\end{equation}
Up to the first order we have
\begin{equation}
\left( {\pounds}_{{\pounds}_{\Delta x}\bar{\varepsilon}}+{\pounds}_{\bar{\varepsilon}}{\pounds}_{\Delta x}\right) \bar{g}=0.
\end{equation}
We further suppose that ${\pounds}_{\Delta x}\bar{\varepsilon}=0$. This means  that our killing vectors are left invariant by our transformation. We should also have ${\pounds}_{\bar{\varepsilon}}{\pounds}_{\Delta x}\bar{g}=0$ which means that the tensor ${\pounds}_{\Delta x}\bar{g}$ is spherically symmetric. To illuminate this more, we will derive explicitly how the metric transforms. The most general metric with spherical symmetry is written as
\begin{equation}
ds^2=g_{00}(t,r)dt^2+2g_{0r}(t,r)dtdr+g(t,r)dr^2+f(t,r)r^2d\Omega^2,
\label{S}
\end{equation}
We have taken the flat FRW as our background metric.
We require that our transformed metric should keep the same form after the transformation  $ x^{u}\rightarrow x^{\mu}+\epsilon^{u}$. Up to the first order, we have

\begin{align}
ds^2=&-dt^2+a^2dr^2+a^2r^2(d \theta^2+\sin(\theta)^2d\varphi^2)\\ \nonumber
=&-dt^2+a^2dr^2+a(t-\epsilon^{0})^2(r-\varepsilon^{r})^2d\Omega^2-2\epsilon^{0}_{,t}dt^2-2\epsilon^{0}_{,i}dx^idt
+a^2(t-\epsilon^{0})dr^2\\&+2a^2(t-\epsilon^{0})\epsilon^{r}_{,t}drdt+2a^2(t-\epsilon^{0})\epsilon^{r}_{,i}dr~dx^i.
\end{align}
where $i$ represents spatial indexes.
We find that the most general transformation is given by
\begin{equation}
\epsilon=(\epsilon^{0}(t,r),\epsilon^{r}(t,r),K(S^2)),
\label{t}
\end{equation}
where $K(S^2)$ are the killing vectors on the sphere.\\

\subsection{Transformation of perturbations}
In this part we derive how perturbations transform under the gauge transformation given by the equation (\ref{t}). We have a perturbed metric as
\begin{align}
ds^2 &= (-a^2+\delta g_{00}(t,r))dt^2+(a^2+\delta g_{rr}(t,r))dr^2+2g_{0r}(t,r)drdt+a^2r^2(1-2E(t,r))d\Omega^2 \\ \nonumber 
&= (-a^2(1-2\mathcal{H}\epsilon^{0})+2a^2\epsilon^{0}_{,0}+\delta g_{00})dt^2+(a^2(1+2\mathcal{H}\epsilon^{0})+2a^2\epsilon^{r}_{,r}+\delta g_{rr})dr^2+\\ & 
2(-a^2\epsilon^{0}_{,r}+ a^2\epsilon^{r}_{,t}+  2g_{0r})dtdr+a^2(1+2\mathcal{H}\epsilon^{0})(r^2+2\epsilon^{r}r-2E)d\Omega^2,
\end{align}
where $\mathcal{H}$ is the conformal Hubble's rate.
 We find that the transformations for perturbed quantities can be written as
\begin{eqnarray}
\label{per-def}
&\delta \tilde{g}_{00}=-2a^2\varphi=\delta g_{00}+2a^2\epsilon^{0}_{,0}+2\mathcal{H}a^2\epsilon^{0},\\\nonumber
&\delta \tilde{g}_{0r}=a^2B=\delta g_{0r}-a^2\epsilon^{r}_{,t}+a^2\epsilon^{0}_{,r},\\\nonumber
&\delta \tilde{g}_{rr}=-2a^2\psi=\delta g_{rr}+2a^2\epsilon^{r}_{,r}+2\mathcal{H}a^2\epsilon^{0},\\\nonumber
&\delta \tilde{g}_{\theta\theta}=-2a^2\tilde{E}=(-2E+2\mathcal{H}a^2\epsilon^{0}+2a^2\frac{\epsilon^{r}}{r})\bar{g}_{\theta\theta}.
\end{eqnarray}
 Finally, the metric perturbations will transform as
\begin{align}
&\delta\varphi=-(\epsilon^{0}_{,0}+\mathcal{H}\epsilon^{0}),\\
&\delta\psi=-\epsilon^{r}_{,r}-\mathcal{H}\epsilon^{0},\\
&\delta B=-\epsilon^{r}_{,t}+\epsilon^{0}_{,r},
\end{align}
\begin{equation}
\delta E=-\frac{\epsilon^{r}}{r}-\mathcal{H}\epsilon^{0}.
\label{metric-trans}
\end{equation}
We know that any scalar $\mathcal{S}$ such as $\delta\rho$ also transforms as
\begin{align}
\delta \mathcal{S}=-(\epsilon^{0}\partial_{t}\bar{\mathcal{S}}+\epsilon^{r}\partial_{r}\bar{\mathcal{S}}).
\end{align}
Expansion for null geodesics (\ref{Ap-expansion}) transforms as
\begin{eqnarray}
\delta\Theta=-(\epsilon^{0}\partial_{t}\bar{\Theta}+\epsilon^{r}\partial_{r}\bar{\Theta}),
\label{tet}
\end{eqnarray}
where $\Theta$ depends on time and space in the background. This equation shows that the expansion is not a gauge invariant quantity and should be used with care.\\

In geometric optic approximation, a null vector is perpendicular to constant phase surfaces which are effectively two dimensional. Consider a light bundle, two neighboring rays are separated by the amount  $\xi^{A}$ in this two dimensional space. The trace of the fractional change matrix $(\frac{d\dot{\xi}^{A}}{d\xi^{B}}) $, where dot denotes covariant derivative along the null direction,  is the expansion. The expansion also quantifies the fractional change of the surface area between neighboring rays. A more rigorous introduction can be found in appendix A.\\

\subsection{Gauge fixing}
Whenever we have perturbed quantities defined on some averaged background in general relativity, we have to define the surface on which this average is taken. As we change the coordinates, the value of these perturbations will change. This has to do with the fact that not all degrees of freedom are physical in perturbation theory. Hence, we face the notion of gauge fixing. One solution is to find quantities that are invariant under coordinate transformations. These gauge invariant quantities are not unique. Some are defined using only metric perturbation as we have for Bardeen potentials, while some may be constructed using metric and matter perturbations. In this section we denote matter density and matter perturbation by $\rho$ and $\delta \rho$ respectively. The matter perturbation will transform as $\delta \tilde{\rho}=\delta\rho-\pounds_{\varepsilon}\bar{\rho}$.

We define two new gauge invariant quantities as
\begin{align}
	\Phi&=\varphi-\frac{1}{a}\frac{\partial}{\partial t}(\frac{a\delta\rho}{\dot{\bar{\rho}}}),\\
	\Psi&=\psi-\frac{\mathcal{H}\delta\rho}{\dot{\bar{\rho}}}+(r(-E+\mathcal{H}\frac{\delta\rho}{\dot{\bar{\rho}}}))_{,r}.
\end{align}
Note that because we constructed these gauge invariant quantities using  matter perturbations and metric perturbations, they do not correspond to Newtonian gauge potentials.
\begin{itemize}
\item
\textbf{Uniform density gauge:} In this gauge we set $\delta\rho=E=0$. By setting $\delta\rho=0$, we fixed the temporal gauge freedom $\varepsilon^{0}$. Using equation (\ref{metric-trans}) we find that spatial gauge freedom can now be fixed by setting $E=0$. The gauge is completely fixed in this case.

The expansion in this gauge is 
\begin{align}
\Theta=\dfrac{\sqrt{2}}{ra}\left(r(1-\varphi)\mathcal{H}+(1+\psi-B) \right). \label{exp-ud}
\end{align}\\

\item
\textbf{Newtonian like gauge:} Another gauge is given by setting $E=B=0$. When we set $E=0$ we have $\epsilon^r=-r\mathcal{H}\epsilon^{0}$. When we set $B=0$ we have
\begin{equation}
\partial_{t}(\mathcal{H}\epsilon^{0})+\frac{\partial_{r}\epsilon^0}{r}=0.
\end{equation}
Let usS introduce the new variables $\mathcal{H}\partial_{t}=\partial_{\eta}$ and $\frac{\partial_{r}}{r}=\partial_{R}$ to solve this equation.
The solution is 
\begin{equation}
A=\mathcal{H}\epsilon^{0}=f(\eta-R),
\label{gaugefreedom}
\end{equation}
where $f(\eta-R)$ is any well behaved function of $(\eta-R)$. In this case the gauge is partially fixed as we have the remaining gauge freedom by equation (\ref{gaugefreedom}). 
In this gauge 
\begin{align} \label{exp-ne}
\Theta=\frac{\sqrt{2}}{ra}\left(\mathcal{H}r(1-\varphi)+1+\psi \right).
\end{align}\\

\end{itemize}

\subsection{Gauge invariant definition for expansion}

Using the equation (\ref{tet}), we define a gauge invariant expansion as 
\begin{align} \label{exp-gi}
\tilde{\Theta}=\Theta-\partial_{t}\bar{\Theta}\frac{\delta\rho}{\dot{\bar{\rho}}}+\partial_{r}\bar{\Theta}(-rE+r\mathcal{H}\frac{\delta\rho}{\dot{\bar{\rho}}}).
\end{align}
This gauge invariant quantity reduces to the expansion in the uniform density gauge where we have $\delta\rho=E=0$. As a result, the expansion in this gauge has a physical meaning and its value is invariant. Since the uniform density gauge has the advantage that all perturbations are captured in the metric, the expansion defined in this gauge is a pure geometric quantity.\\

The expansion for outgoing null geodesics in the flat FRW background is 
\begin{align}
\Theta=\frac{\sqrt{2}}{ra}\left(\mathcal{H}r+1 \right).
\end{align}\\
Note that this quantity never vanishes for out going null geodesics in the background. It is crucial to note that expansion does vanish for ingoing  null geodesics at $r_{cos}=\frac{1}{\mathcal{H}}$. As described in the Appendix A, this surface is the cosmological horizon, while a black hole is defined by vanishing outgoing expansion. In other words a homogeneous and isotropic universe never forms a black hole regardless of its density.\\

We can use this gauge invariant expansion to answer our main point  that linear perturbations do not  result in black hole formation.  The equations (\ref{exp-gi}),  (\ref{exp-ud}) and  (\ref{exp-ne}) reveal that as long as we have $|g_{ab}| \gg |\delta g_{ab}|$  the outgoing expansion is
 nonzero. As a result, we  do not have black holes inside the cosmological horizon in the linear regime. 
 This is in contrast to the intuition that in early universe when density is high, small perturbations lead to black holes.

\subsection{PBH  formation in the long wavelength limit}
One interesting scenario is that perturbations collapse to black holes when they are already outside the horizon. We argue this can not occur.  Although to study this scenario the linear perturbation theory we used in the last section can be applied here, if the superhorizon perturbations are linear, the superhorizon limit can be studied in a different way. The superhorizon limit has the advantage that the full non-linear equations can be simplified without assuming that potential perturbations are small. To study this scenario we need the long wavelength limit of full non-linear equations. The method, named gradient expansion, has been proved useful in studying the early universe models \cite{Lyth:2004gb}. In this method each quantity is expanded in powers of $\epsilon=\frac{k}{\mathcal{H}}$. This method has been extended to PBHs to set their initial conditions outside the horizon \cite{Shibata:1999zs}. We just give the main points here.\\

Before taking the long wave length limit, the metric is written in a conformally decomposed form as
\begin{align}
ds^2=-(\alpha^2-\psi^4\beta^2r^2)dt^2+2\psi^4a^2\beta rdrdt+\psi^4a^2(dr^2+r^2d\Omega^2),
\end{align} 
where $a$ is the scale factor. The extrinsic curvature is also decomposed to trace and traceless parts as $K_{ij}=\psi^4a^2\tilde{A}_{ij}+\frac{\psi^4a^2\tilde{\gamma}_{ij}K}{3}$ where $\tilde{\gamma_{ij}}$ is the spatial flat metric in spherical coordinates. Expansion in this metric is given by
\begin{align}
\Theta=2\frac{\dot{a}}{a}+A+\frac{1}{\psi^2a}(\frac{2}{r}+\frac{4\partial_{r}\psi}{\psi}),
\end{align}
and $A=\tilde{A}^{r}_{r}$. To know the value of the expansion on superhorizon scales we will need to estimate the order of magnitude for $\psi$ and  its first derivative $\partial_{r}\psi$. In the gradient expansion, it is assumed that $\delta \rho=\mathcal{O}(\epsilon^{2})$ and the gradient of any variable is $\partial_{i}\psi=\psi\times \mathcal{O}(\epsilon)$. The long wave length limit of other variables is found using the $3+1$ formalism equations. This assumption leads to $\psi-1\sim {\mathcal O}(\epsilon^{0}),\partial_{r}\psi\sim {\mathcal O}(\epsilon^{1})$ \cite{Harada:2015yda, Shibata:1999zs}. It is important to note that here $\psi$ is not necessary small. It is found that the first term is positive and the third term in negligible. As a result, a black hole will not form as long as  the perturbations are outside the Hubble horizon.\\

\section{Pbh formation in non-linear regime: Spherical collapse in radiation era }

We showed that black holes do not form in the linear regime and superhorizon scales. Now we consider the non-linear regime. While, the non-linear study of black-hole formation in radiation era can be done in numerical relativity, analytical models can  help to illuminate our physical intuition.  
To study black hole formation in non-linear regime we consider a toy model. We take the spherical collapse of an over-dense sphere in a flat FRW radiation-filled background. This region is modeled by a closed FRW metric. We also include velocity perturbation in our model in terms of different Hubble's expansion rates. We show that the overdensity of the collapsing region is non-linear when the apparent horizon forms and when initial density perturbations are small.

By demanding that our initial conditions match the perturbation theory we find a constraint on the value of $\delta^{h}_{0}$ which is the initial Hubble' rate perturbation. The value of the overdensity at horizon entry is found by demanding to have the sound horizon inside the cosmological horizon. This results in $\delta_{0}=0.7$. Since the value of $\delta_{0}$ is related to the primordial power spectrum which is set by inflation, the inflationary perturbations should have substantial power at that scale to form primordial black holes.\\

Adopting the spherical collapse model for the PBH formation, we assign a spherically averaged overdensity  $\delta(R,t)$ to every over-dense patch. The smoothed overdensity can be found via applying an appropriate window function. Generally, the window function picks up only long modes with wavelength larger than the smoothing scale. This can be justified as the (spatial) oscillations of the short modes $q \gtrsim 1/R $ are averaged out on the smoothing scale $R$. Hence, we assume the following uniform overdensity $\delta_{\mathrm{sc}}$ for the fluctuations    
\begin{align}
	\delta_{\mathrm{sc}}= \int d^3 \q \, \delta(q) \vert W(q R)\vert^2.
\end{align}
This is roughly equal to $\delta_{\mathrm{sc}}\approx\sqrt{\mathcal{P}(k)}|_{k\sim\frac{1}{R}}=\sqrt{\frac{k^3}{2\pi}P(k)}|_{k\sim\frac{1}{R}}$. That is our amplitude is related to the square root of the primordial power spectrum evaluated at the scale $R$.\\

According to the Birkhoff theorem, the geometry of a spherically symmetric homogeneous over-dense region is well described by a FRW metric
\begin{equation}
ds^2=-dt^2+R(t)^2(d\chi^2+\sin^2(\chi)d\Omega^2),
\end{equation}
 with a density $\rho= \bar{\rho} (1+\delta_0)$ where $\delta_0$ is the initial density contrast and  $K$ is the curvature constant.
The Friedmann equation for this region is given by
\begin{align}
\tilde{H}^{2}=\frac{D}{R^4}-\frac{K}{R^2},
\label{Friedman}
\end{align}
where $D = 8\pi G \bar{\rho}(1+\delta_0) R_0^4 /3$ quantifies the total initial energy of the over-dense region. Using the apparent horizon definition in the Appendix A, it can be shown that the apparent horizon is located at
\be
\dot{R}=\cot(\chi).
\ee
In the case that the black hole boundary is the $\chi=\frac{\pi}{2}$ surface, the apparent horizon or the black hole boundary will be located at the turnaround point, where we have $\dot{R}=0$.

Using the equation (\ref{Friedman}), we will obtain the scale factor of the collapsing sphere $R(t)$ as 
\begin{equation}
\label{R-t}
R(t)=\sqrt{R_0^2+ \sqrt{4D-4KR_{0}^2} \,(t-t_{0}) -K
	(t-t_{0})^2}.
\end{equation}
The over-dense region first expands till the time of maximum extension $t_{m}$ and collapses afterwards, which is named as turnaround. The maximum extension $R_m$ is found to be
\begin{equation}
R_{m}=\sqrt{\frac{D}{K}}.
\end{equation}
The turnaround time is
\begin{equation}
t_{m}=\frac{\sqrt{D-K {R_{0}}^2}}{K}+{t_{0}}.
\end{equation}

In order to find the exact time dependence of the radius of collapsing sphere, one has to determine the effective curvature of over-dense region. To this end, the initial condition for this problem can be fixed at early times well before  $t \ll t_m$. The linear theory predicts that at very early times, the overdensity goes as $\delta \rho(t) \sim a(t)$ where $a(t)= \sqrt[4]{4A}~t^\frac{1}{2}$ is the scale factor of the FRW background .

Let us assume, without any reference to the linear theory, the following initial condition for the radial velocity of the over-dense sphere
\begin{align}
\dfrac{\dot{R}(t_0)}{R(t_0)} \equiv {\tilde{H}}_0 = H_0 (1+\delta^h_0),
\end{align}
where $H_0$ denotes the Hubble parameter of the background metric and $\delta_{0}^{h}$ is the initial Hubble parameter perturbation. Using the Friedmann equation \eqref{Friedman}, this initial condition totally determines the curvature constant of the interior metric as
\begin{align}
K = R_0^2 ( D/R_0^4 -{\tilde{H}}_0^2 ) = R_0^2 H_0^2 \left[(1+\delta_0)-(1+\delta^h_0)^2\right].
\end{align} 
Using the above relation, the overdensity at turnaround is found to be
\begin{align}
\delta(t_m) = \dfrac{\rho_{\mathrm{sc}}-\bar{\rho}}{\bar{\rho}} \Big\vert_{t_m} = 3 + {\delta_{o}}^{2}-4\,{\delta_0^h}^2 + {\cal O}(\delta^3).
\end{align}
This interesting result shows that the overdensity at turnaround has the universal value of $3 $ up to corrections of the second order of the initial perturbations. This result emphasizes that at turnaround $t=t_m$, when the apparent horizon forms and collapse starts, the overdensity is already non-linear.\\

The next step would be specifying the initial condition of the spherical collapse. At early times well before the curvature term dominates and the dynamics of the over-dense region leads to recollapse, the evolution of the smoothed patch must coincide with the linear perturbation theory. Specifically the prediction of the spherical collapse model for the evolution of an overdensity must match the evolution of an overdensity in the linear theory
\begin{align}
\delta(t) \sim a^2(t).
\end{align}
Expanding \eqref{R-t} for early times to linear order of the perturbations and noting that $a(t)= \sqrt[4]{4A}~t^\frac{1}{2}$, overdensity of the collapsing region is derived as
\begin{align}
\delta=(\delta_{0}-2\delta^h_0)\,H_{0}t+\frac{\delta_{0}+2\delta^h_0}{4H_{0}t}.
\end{align}
As time elapses, the decaying solution decays and we get 
\begin{align}
\delta=(\delta_{0}/2-\delta^h_0)~t/t_0,
\end{align}
in which we have used the the fact that in the radiation dominated universe one has, $H(t)= 1/2t$. By trying to match this result with what we expect from the linear theory, we  find a very important constraint for the velocity perturbation as
\begin{align}
\delta^h_0 = - \delta_{0}/2,
\end{align}
That is if the spherical collapse model is to be considered a linear perturbation, its velocity perturbation must be minus half of its density perturbation.\\
  
\begin{figure}[htbp!]
\centering
\includegraphics[width=0.5 \columnwidth]{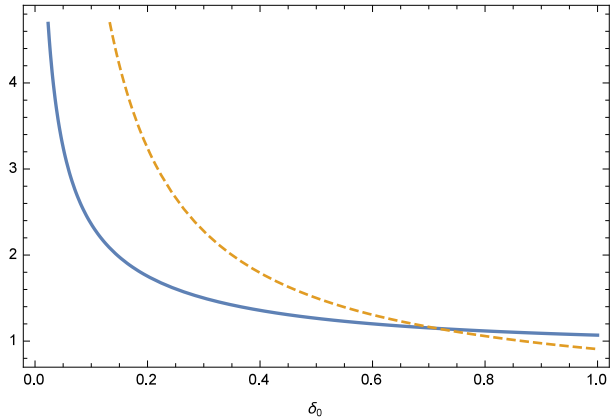}
\caption{The physical sound horizon  $L_{j}$ and $R_{max}=\frac{\pi}{2}R_{m}$. Allowed region is where sound horizon is inside the apparent horizon. The dashed line and the thick line represent the sound horizon and the apparent horizon respectively.}
\label{fig:1}	
\end{figure}

Let us now set our initial condition when the central region enters the background horizon $a_{0}=\frac{2}{\pi H}$ and $t_{0}=\frac{1}{2H_{0}}$ and use the initial condition on $\delta^h_0$ which we obtained.
The behavior of a medium to collapse under its own gravity is determined by a length scale, named  Jeans length, where pressure gradients tend to oppose the collapse. The Jeans length for a fluid with sound speed $c_{s}$ is given by $l_{j}=c_{s}\sqrt{\frac{\pi}{G\rho}}$. However, this scale is derived using Newtonian considerations in the perturbation theory. Since we study black holes in radiation dominated era, we will need a relativistic definition. One could say that the physical length when horizon forms, turnaround time in our model, should be less than the sound horizon \cite{Harada:2013epa}. The sound horizon is defined by
\begin{equation}
L_{j}=\frac{1}{\sqrt{3}}a(t_{m})\int_{t_{0}}^{t_{m}}\frac{dt'}{a(t')}.
\end{equation}
The figure (\ref{fig:1}) shows the physical sound horizon and physical turnaround radius for different $\delta_{0}$.
We find that in the case of perturbative initial conditions, the threshold value of density perturbation at horizon entry must be larger than $0.7$ to overcome the pressure gradients. This requires that the initial perturbations have enough power at the entry scale. However, this is not the case with simple models of inflation. Note that pasting the initial conditions to the perturbation theory solution is different than taking uniform Hubble slicing as implemented in \cite{Harada:2013epa}. By pasting our solution we are mildly taking in to account the pressure gradients initially. Hence, we have larger threshold. \\

\section{Conclusion}

In this paper our aim is twofold. We apply the concept of apparent horizon for dynamical black holes to revisit the formation of primordial black holes  in the early universe for both linear and non-linear regimes. Given that the event horizon is limited by the global nature of spacetime evolution, it is advantageous to define a black hole by its apparent horizon. The apparent horizon is quantified by the expansion of null geodesics which distinguishes the trapped surface in the black hole case.\\
 
First, we develop the perturbation theory in the spherically symmetric spacetime and then we fix two gauges. We also define a gauge invariant quantity for the expansion of null geodesics. We have shown that it is not possible to have  trapped surfaces in linear and superhorizon regimes. Hence, the primordial black holes cannot form in these cases.  One may think that in the early universe where density is very high, any perturbation  can lead to a black  hole. But this work shows that  in the linear regime the black holes can not form even when we have high densities. \\

In non-linear regime, we implemented a closed FRW to model the collapse of primordial black holes in the radiation dominated era. In our model the boundary of the collapsing region is located at $\chi=\pi/2$. The turnaround point coincides with the black hole apparent horizon.   We use the initial condition from the perturbation theory as we want our model to be linear initially. Our approach allows to have density and velocity perturbations. Most important results are that this model gives a constraint for the matching condition and the primordial black hole's overdensity at the onset of the black hole formation has the universal value $\delta > 3$. This verifies our result  that the Primordial black holes will form in the non-linear regime.  Applying the sound horizon constraint, we have shown that the threshold value of density perturbations at horizon re-entry must be larger than $\delta _{th} >  0.7$ to overcome the pressure gradients.\\

{\bf Acknowledgments:}
\\

We would like to thank John C. Miller, Misao Sasaki, Ilia Musco and Hassan Firouzjahi for useful
comments.
\\

\appendix
\section{Light cone dynamics}

Let $S$ be a closed and orientable two-surface which is (smoothly)
embedded in a four-dimensional time-oriented spacetime $(M,g_{ab})$ which has a metric compatible covariant derivative $\nabla_a$.  There are just two future-directed null directions normal to $S$. Let $\ell^a$
and $n^a$ be null vector fields pointing in these directions; in situations where this is meaningful we will always take $\ell^a$ and $n^a$ as  outgoing and ingoing respectively.  If we further suppose  that $\ell \cdot n = -1$ then there is only one remaining degree
of rescaling freedom in the definition of these vector fields. 

The intrinsic geometry of $S$ is defined by the induced metric. The definition of these quantities is independent of the choice of null vectors made above. Nevertheless, for our purposes it is most useful to express them in terms of these vectors. Hence, the induced metric on
$S$ can be written as 
\ba \label{metric2}
q_{ab} = g_{ab} + \ell_a n_b + \ell_b n_a \label{q} .
\ea 
The covariant derivative operator ${d}_a=q_a^b \nabla_b$ and
(two-dimensional) Ricci scalar $\tilde{R}$ are defined on this two-surface. The extrinsic geometry which shows how $S$ is embedded in $M$. The extrinsic curvature  is defined by how the a normal vectors
change over $S$ as in the usual way. The extrinsic curvatures are
\be
k^{(\ell)}_{ab} = q_a^c q_b^d \nabla_c \ell_d 
\quad  \mbox{and}  \quad
k^{(n)}_{ab} = q_a^c q_b^d \nabla_c n_d  \, . \label{kln}
\ee
We can decompose the extrinsic curvatures as
\ba
k^{(\ell)}_{ab} = \frac{1}{2} \theta_\ell q_{ab} + \sigma^{(\ell)}_{ab} 
\quad \mbox{and} \quad 
k^{(n)}_ {ab} = \frac{1}{2} \Theta_n q_{ab} + \sigma^{(n)}_{ab} \, . \label{dq}
\ea 
where
\begin{eqnarray} \label{Ap-expansion} \Theta_{(\ell)} = q^{ab} \nabla_{a} \ell_{b} 
\quad \mbox{ and } \quad 
\Theta_{(n)} = q^{ab} \nabla_{a} n_{b} \, , 
\end{eqnarray} 
are  the \emph{expansions} which are the traces of the extrinsic curvatures and the \emph{shears} 
\begin{equation}\label{shears}
\sigma^{(\ell)}_{ab} \equiv \left(q_{(a}^c q_{b)}^d 
- \frac{1}{2} q_{ab} q^{cd} \right) \nabla_{c} \ell_{d}  
\quad \mbox{and} \quad 
\sigma^{(n)}_{ab} \equiv \left(q_{(a}^c q_{b)}^d - \frac{1}{2} q_{ab} q^{cd}
\right) \nabla_{c} n_{d}  \, ,
\end{equation}
are the trace-free parts. The rotation tensors are
\begin{equation}\label{shears}
w^{(\ell)}_{ab} \equiv q_{[a}^c q_{b]}^d \nabla_{c} \ell_{d}  
\quad \mbox{and} \quad 
w^{(n)}_{ab} \equiv q_{[a}^c q_{b]}^d \nabla_{c} n_{d}.
\end{equation}
Here $()$ and $[~]$ denote symmetrization and antisymmetrization of indexes.
For hypersurface orthogonal null foliation  \cite{poisson-book} the rotation tensor is zero.  If we define 
\be
\kappa_X= -n_a X^b \nabla_{b} \ell^{a},
\ee
the Raychaudhuri equation will be
\begin{equation}\label{Ray}  
\pounds_\ell \Theta_{(\ell)} = \kappa_\ell \Theta_{(\ell)} -  (1/2) \Theta_{(\ell)}^{2} 
- \sigma_{ab}^{(\ell)} \sigma^{(\ell)ab}
- G_{ab} \ell^{a} \ell^{b}
\, \, .  
\end{equation}
In the case of the affine parameter for $\ell$, we find that the $\kappa_\ell=0$. Similar equation for null geodesic is 

\begin{equation}\label{Ray2}  
\pounds_n \Theta_{(\ell)} = -\tilde{R}/2 +w_a w^a -d_a w^a + G_{ab} \ell^{a} n^{b}
,  
\end{equation}
where is $w_a=-q^b_a n_c \nabla_b \ell^c$. In the spherically symmetric case
\be
ds^2=- e^{\nu(t,r)} dt^2 +e^{\psi(t,r)} dr^2+R(t,r)^2 d\Omega^2,
\ee
and shear free foliation, these two equations reduce to
\ba\label{Ray}  
\pounds_\ell \Theta_{(\ell)} = \kappa_\ell \Theta_{(\ell)} -  (1/2) \Theta_{(\ell)}^{2} - G_{ab} \ell^{a} \ell^{b}, \nonumber \\
\pounds_n \Theta_{(\ell)} = -\frac{1}{ R^2} +\Theta_\ell \Theta_n + G_{ab} \ell^{a} n^{b}
.
\ea
We can define the marginally trapped surface $\bar{H}  $ as $\Theta_\ell =0$ and this surface is foliated by spacelike two spheres. We can always write a tangent vector to $\bar{H} $ as
\be
V^a=\ell^a-C n^a
\ee
Since on the $\bar{H} $ we have $\pounds_V \Theta_{(\ell)}=0$ we find that
\be
C=\frac{\pounds_\ell \Theta_{(\ell)}}{\pounds_n \Theta_{(\ell)}}|_{H}= \frac{ G_{ab} \ell^{a} \ell^{b} }{\frac{1}{ R^2} - G_{ab} \ell^{a} n^{b}}.
\ee

\textbf{Black hole definition}: A smooth, three-dimensional, space-like sub-manifold (possibly with boundary) $\bar{H}  $ of spacetime is said to be a trapping horizon  if it can be foliated by a family of closed 2-manifolds such that  on each leaf $S$ the expansion $\Theta_{(\ell)}$ of one null normal $\ell^\mu$  vanishes; and the expansion $\Theta_{(n)}<0$ of the other null normal $n^\mu$ is negative. This surface separates the trapped surface, $\Theta_{(n)}, \Theta_{(\ell)}<0$, from untrapped one $\Theta_{(n)}<0,~ \Theta_{(\ell)}>0$. \\

There are diverse definition for the black hole boundary in the dynamical cases. One definition is the dynamical horizon as the black hole boundary \cite{Ashtekar:2004cn} which is a spacelike trapping horizon.  A foliation independent definition for the apparent horizon comes from trapping boundary which is boundary of all trapped surfaces.\\

Similarly, the cosmological horizon is defined as the  closed 2-manifolds such that  on each leaf $S$ the expansion $\Theta_{(n)}=0$ of one null normal $n^\mu$  vanishes; and the expansion $\Theta_{(\ell)} > 0$ of the other null normal $\ell^\mu$ is positive. \\

\section{Spherical symmetric perturbation in terms of spherical harmonics}
Perturbations of spherically symmetric spacetimes are best studied in $2+2$ decomposition. The perturbations are expanded in spherical harmonic bases and are decomposed under parity transformation to odd parity and even parity. Gauge invariant expressions can be constructed for each case. But $l=0,1$ should be studied separately. In this section we will review perturbations of spherically symmetric spacetimes \cite{Gundlach:1999bt}.\\

We have a background metric which we write as
\begin{equation}
ds^2=g_{AB}dx^{A}dx^{B}+r(x^{A})^2 h_{ab}dx^adx^b.
\end{equation} 
Scalar perturbations are expanded in $Y^{l}_{m}$. Vector bases are constructed for even parity as
\begin{eqnarray}
Y^{l}_{m,a},
\end{eqnarray}
and for odd parity as
\begin{eqnarray}
 S^{l}_{m}=\varepsilon^{b}_{a}Y^{l}_{m,b},
\end{eqnarray}
where $\varepsilon _{ab}$ is the antisymmetric  Levi-Civita tensor and lower Latin indexes refer to polar angels. Even parity tensors are given by
\begin{eqnarray}
Y^{l}_{m}h_{ab},\\\nonumber
 Z^{l}_{m}=Y^{l}_{m:ab}+\frac{l(l+1)}{2}Y^{l}_{m}h_{ab},
\end{eqnarray} and odd parity tensors by
\begin{eqnarray}
S^{l}_{m:ab}+S^{l}_{m:ba},
\end{eqnarray}
In linear regime $Y^{l}_{m}$ with different $l,m$ are decoupled. Even and odd parity perturbations are also decoupled.

A gauge transformation could be given by $\varepsilon_{\mu}=(\zeta Y^{l}_{m},r^2\varepsilon Y^{l}_{m,a}+r^2MS_{a})$.
We have three degree of freedom for odd parity perturbations and one odd parity degree of freedom from gauge transformations leaving two odd parity gauge invariant perturbations. We can remove three degrees of freedom by gauge transformations for odd parity perturbations. We are left with four even parity gauge invariant perturbations. 
We write the metric in the case of odd parity perturbations as
\begin{equation}
g^{odd}_{\mu\nu}=
\begin{pmatrix}
0& h_{A}S_{a}\\
h_{A}S_a & h(S_{a:b}+S_{b:a})
\end{pmatrix},
\end{equation}

and in the case of even parity perturbations as
$$
g^{even}_{\mu\nu}=\begin{pmatrix}
h_{AB}Y^{l}_{m}& H_{A}Y^{l}_{m,a}\\
H_{A}Y^{l}_{m,a}& r^2(KY^{l}_{m}h_{ab}+GY^{l}_{m:ab})
\end{pmatrix}
$$ 

\subsection{gauge invariant perturbations}
The even parity gauge invariant perturbations are given by
\begin{eqnarray}
k_{AB}=&h_{AB}-(p_{A|B}+p_{B|A})\\\nonumber
k=&K-2v^Ap_{A}
\end{eqnarray}
where $v_A=\frac{r|A}{r}$ and $p_{A}=H_{A}-\frac{1}{2}r^2G_{|A}$. The odd parity perturbations are
\begin{eqnarray}
k_{A}=h_{A}-r^2(\frac{h}{r^2})_{|A}.
\end{eqnarray}
The gauge in which $h=H_{A}=G=0$ is called Regge-Wheeler gauge.
\section{Einstein's equations}

In this section we write the Einstein equations for spherical perturbations of a flat FRW universe with a prefect fluid with the equation of state $P=\omega\rho$. The metric is presented by the equation (\ref{per-def}). In Newtonian-like gauge perturbation equations are
\begin{align}
&\delta G_{00}=\frac{2 a'(t) \partial_{t}\psi }{a(t)}+\frac{2 \psi }{r^2}+\frac{2 \partial_r\psi }{r}\\\nonumber
&\delta G_{01}=\frac{2 \partial_{t}\psi}{r}-\frac{2 a'(t) \partial_r\Phi }{a(t)}\\
&\delta G_{11}=-\frac{4 a''(t) \Phi }{a(t)}-\frac{4 a''(t) \psi }{a(t)}-\frac{2 a'(t) \partial_{t}\Phi }{a(t)}+\frac{2 a'(t)^2 \Phi
	}{a(t)^2}+\frac{2 a'(t)^2 \psi }{a(t)^2}-\frac{2 \psi }{r^2}\\\nonumber
&-\frac{2 \partial_{r}\Phi }{r}\\
\delta G_{22}&=-\frac{4 r^2 a''(t) \Phi }{a(t)}-\frac{2 r^2 a'(t) \partial_{t}\Phi }{a(t)}+\frac{2 r^2 a'(t)^2 \Phi }{a(t)^2}-\frac{2 r^2 a'(t) \partial_{t}\psi}{a(t)}-r^2\partial^2_{r}\Phi \\ \nonumber
-r^2 \partial^2_{t}\psi -r \partial_{r}\Phi-r \partial_{r}\psi
\end{align}
In the uniform density gauge  we have
\begin{align}
G_{00}&=\frac{2 \left(r^2 a'(t) B^{(0,1)}(t,r)+2 r a'(t) B(t,r)+r^2 a'(t) \psi ^{(1,0)}(t,r)+r a(t) \psi ^{(0,1)}(t,r)+a(t) \psi (t,r)\right)}{r^2 a(t)}\\
G_{01}&=\frac{2 a''(t) B(t,r)}{a(t)}-\frac{a'(t)^2 B(t,r)}{a(t)^2}-\frac{2 a'(t) \Phi ^{(0,1)}(t,r)}{a(t)}+\frac{2 \psi ^{(1,0)}(t,r)}{r}\\
G_{11}&=-\frac{4 a''(t) \Phi (t,r)}{a(t)}-\frac{4 a''(t) \psi (t,r)}{a(t)}-\frac{4 a'(t) B(t,r)}{r a(t)}-\frac{2 a'(t) \Phi ^{(1,0)}(t,r)}{a(t)}+\frac{2
	a'(t)^2 \Phi (t,r)}{a(t)^2}+\\
&\frac{2 a'(t)^2 \psi (t,r)}{a(t)^2}-\frac{2 B^{(1,0)}(t,r)}{r}-\frac{2 \psi (t,r)}{r^2}-\frac{2 \Phi
	^{(0,1)}(t,r)}{r}\\
G_{22}&=-\frac{4 r^2 a''(t) \Phi (t,r)}{a(t)}-\frac{2 r^2 a'(t) B^{(0,1)}(t,r)}{a(t)}-\frac{2 r a'(t) B(t,r)}{a(t)}-\frac{2 r^2 a'(t) \Phi
	^{(1,0)}(t,r)}{a(t)}\\
&+\frac{2 r^2 a'(t)^2 \Phi (t,r)}{a(t)^2}-\frac{2 r^2 a'(t) \psi ^{(1,0)}(t,r)}{a(t)}-r^2 B^{(1,1)}(t,r)-r B^{(1,0)}(t,r)-r^2
\Phi ^{(0,2)}(t,r)\\
&-r^2 \psi ^{(2,0)}(t,r)-r \Phi ^{(0,1)}(t,r)-r \psi ^{(0,1)}(t,r),
\end{align}
where $(i,j)$ means $i^{th}$ derivative with respect to the first argument and $j^{th}$ derivative with respect to the second argument. 

Perturbations of the prefect fluid in a general gauge are given by
\begin{align}
&\delta T_{00}= a^2\delta\rho+2\rho a^2\phi,\\
&\delta T_{01}=a(1+\omega)\bar{\rho}\delta u_{1}+a^2\omega B\bar{\rho},\\
&\delta T_{11}=-2a^2\omega\bar{\rho}\psi+a^2\omega\delta\rho,\\
&\delta T_{ab}=\omega\bar{\rho}\delta g_{ab}+\omega \bar{g}_{ab}\delta\rho,
\end{align}
where $\delta u_{1}$ is the radial velocity perturbation.


\begin{thebibliography}{99}

\bibitem{Carr:1975qj} 
B.~J.~Carr,
Astrophys.\ J.\  {\bf 201}, 1 (1975).



\bibitem{Niemeyer:1999ak} 
J.~C.~Niemeyer and K.~Jedamzik,
Phys.\ Rev.\ D {\bf 59}, 124013 (1999).

\bibitem{Musco:2004ak} 
I.~Musco, J.~C.~Miller and L.~Rezzolla,
Class.\ Quant.\ Grav.\  {\bf 22}, 1405 (2005).


\bibitem{Shibata:1999zs} 
M.~Shibata and M.~Sasaki,
Phys.\ Rev.\ D {\bf 60}, 084002 (1999).


\bibitem{Young:2014ana} 
A.~G.~Polnarev and I.~Musco,
Class.\ Quant.\ Grav.\  {\bf 24}, 1405 (2007); S.~Young, C.~T.~Byrnes and M.~Sasaki,
JCAP {\bf 1407}, 045 (2014).

\bibitem{new-thre} 
 A.~G.~Polnarev and I.~Musco,
 Class.\ Quant.\ Grav.\  {\bf 24}, 1405 (2007); I.~Musco and J.~C.~Miller,
Class.\ Quant.\ Grav.\  {\bf 30}, 145009 (2013); J.~Bloomfield, D.~Bulhosa and S.~Face,
arXiv:1504.02071 [gr-qc].


\bibitem{Page:1976wx} 
D.~N.~Page and S.~W.~Hawking,
Astrophys.\ J.\  {\bf 206}, 1 (1976); Ukwatta, T. N., et al. 
"Observational Characteristics of the Final Stages of Evaporating Primordial Black Holes." arXiv preprint arXiv:1507.01648 (2015).


\bibitem{Carr:2009jm} 
B.~J.~Carr, K.~Kohri, Y.~Sendouda and J.~Yokoyama,
Phys.\ Rev.\ D {\bf 81}, 104019 (2010).


\bibitem{Ashtekar:2004cn} 
A.~Ashtekar and B.~Krishnan,
Living Rev.\ Rel.\  {\bf 7}, 10 (2004).

\bibitem{Firouzjaee:2010ia} 
J.~T.~Firouzjaee, M.~P.~Mood and R.~Mansouri,
Gen.\ Rel.\ Grav.\  {\bf 44}, 639 (2012).


\bibitem{cosmological black hole}
Faraoni, Valerio. "Horizons." Cosmological and Black Hole Apparent Horizons. Springer International Publishing, 2015. 25-57; A. Krasinski and C. Hellaby, Phys. Rev. D 69, 023502 (2004); J.~T.~Firouzjaee and R.~Mansouri, Gen.\ Rel.\ Grav.\  {\bf 42}, 2431 (2010); R.~Moradi, Javad~T.~Firouzjaee and R.~Mansouri, Class.\ Quant.\ Grav.\  {\bf 32}, no. 21, 215001 (2015).


\bibitem{tunnelingbh} 
R.~Di Criscienzo, S.~A.~Hayward, M.~Nadalini, L.~Vanzo and S.~Zerbini,
Class.\ Quant.\ Grav.\  {\bf 27}, 015006 (2010); J.~T.~Firouzjaee and R.~Mansouri,
Europhys.\ Lett.\  {\bf 97}, 29002 (2012).

\bibitem{firouzjaee-ellis14} 
J.~T.~Firouzjaee and G.~F.~R.~Ellis,
Gen.\ Rel.\ Grav.\  {\bf 47}, no. 2, 6 (2015);
J.~T.~Firouzjaee and G.~F.~R.~Ellis,
arXiv:1511.04316 [gr-qc].

\bibitem{firouzjaee-penn} 
J.~T.~Firouzjaee,
Int.\ J.\ Mod.\ Phys.\ D {\bf 21}, 1250039 (2012).


\bibitem{hofman-2011} 
M.~Kopp, S.~Hofmann and J.~Weller,
Phys.\ Rev.\ D {\bf 83}, 124025 (2011).


\bibitem{Harada:2013epa} 
T.~Harada, C.~M.~Yoo and K.~Kohri,
Phys.\ Rev.\ D {\bf 88}, no. 8, 084051 (2013)
Erratum: [Phys.\ Rev.\ D {\bf 89}, no. 2, 029903 (2014)]
doi:10.1103/PhysRevD.88.084051, 10.1103/PhysRevD.89.029903.


\bibitem{Harada:2015yda} 
T.~Harada, C.~M.~Yoo, T.~Nakama and Y.~Koga,
Phys.\ Rev.\ D {\bf 91}, no. 8, 084057 (2015).


\bibitem{Gundlach:1999bt} 
C.~Gundlach and J.~M.~Martin-Garcia,
Phys.\ Rev.\ D {\bf 61}, 084024 (2000).

\bibitem{poisson-book} 
Poisson, Eric. A relativist's toolkit: the mathematics of black-hole mechanics. Cambridge university press, 2004.

\bibitem{Lyth:2004gb} 
D.~H.~Lyth, K.~A.~Malik and M.~Sasaki,
JCAP {\bf 0505}, 004 (2005).


	
\end{thebibliography}
\end{document}